\documentstyle[12pt]{article}
\newcommand{\be}{\begin{equation}}
\newcommand{\ee}{\end{equation}}
\newcommand{\rref}[1]{(\ref{#1})}

\begin{document}

\begin{flushright}
ULB--TH 01/96 \\
hep-th/9601074 \\
January 1996\\
(revised May 1996)
\end{flushright}

\vspace{.8cm}

\begin{center}
{\LARGE \bf 2+1 Black Hole with a Dynamical}\\
{\LARGE \bf Conical Singularity}\\
\vspace{1cm}

{\large Riccardo Argurio}\footnote{
Aspirant du Fonds National de la Recherche Scientifique; e-mail:
rargurio@ulb.ac.be} \addtocounter{footnote}{-1}\\
\vspace{.4cm}
{\it Service de Physique Th\'eorique}\\
{\it Universit\'e Libre de Bruxelles, Campus Plaine, C.P.225}\\
{\it Boulevard du Triomphe, B-1050 Bruxelles, Belgium}
\end{center}

\vspace{1cm}

\vspace{1cm}

\begin{abstract}
We find black hole solutions to Euclidean 2+1 gravity coupled to a relativistic
particle which have a dynamical conical singularity at the horizon.
These solutions mimic the tree level contribution to the partition function
of gravity coupled to a quantum field theory. They are found to
violate the standard area law for black hole
entropy, their entropy being proportional to the total opening angle.
Since each solution depends on the number of windings of the particle path
around the horizon, the significance of their summation
in the path integral is considered.

\end{abstract}

PACS numbers: 04.70.-s, 04.60.Kz, 05.70.Ce.

\newpage

\section{Introduction}

The origin of the entropy of black holes is a problem which focused
a large amount of work, ever since the formulation of the laws of black
hole thermodynamics \cite{4lws}. 
Despite the difficulties raised by the
question of what the statistical mechanical interpretation of black
hole entropy is, it is well known how to compute it by thermodynamical
techniques.
The entropy arises from the tree level contribution
of the Euclidean gravitational action to the partition 
function \cite{gh}, and is essentially due to the presence of a horizon.
It bears no direct relation to the fluctuations of any external (quantum) field
and is therefore an intrinsic characteristic of the black hole.

Thus far, there have been several attempts to give an interpretation of 
black hole entropy in terms of microscopic degrees of freedom. Most
recently, promising results have come from string theory 
\cite{dbranes} where counting
of elementary and solitonic states agrees with the area law 
for black hole entropy (which can be generalized to any dimension):
\be S=\frac{A_h}{4G}, \label{arealaw}\ee
where $A_h$ is the (generalized) area of the horizon and $G$ is the
Newton constant.
However, string theory does not provide yet a general explanation of 
this formula; rather, it has to be checked case by case, and indeed it
has been checked only in a few special cases.
In a different direction, and in the context of 2+1 dimensional
black holes, there was also the recent proposal of Ref.~\cite{carlip}.

Yet another direction is to consider examples for which the area law
\rref{arealaw} is corrected or violated .
A celebrated example where such a violation
occurs is the extremal black hole which, according to
Refs.~\cite{hhr,teit}, has vanishing entropy (see, however, Ref.~\cite{horo}
for a string theory-based confutation of this claim).
In a recent work, Englert, Houart and Windey \cite{ehw} have
found another interesting example:
a dynamical conical singularity produced by a Nambu-Goto
string wound around the horizon decreases the entropy of the black hole
proportionally to the deficit angle. In Refs.~\cite{cpw,dgt} a similar
mechanism was used to shift the free energy of the black hole in models
in which the string action
arises as the effective action of a vortex in a spontaneously broken gauge
theory.

In this paper we will focus on 2+1 dimensional anti-de Sitter black holes
\cite{bhtz}. The reason is twofold: 2+1 dimensional gravity is simpler
to handle than in 3+1 dimensions, especially when one is dealing with
matter consisting of point particles; BTZ black holes have a nicer 
thermodynamical behaviour than Schwarzschild ones. This is seen as 
follows. 
Schwarzschild black holes have a negative 
heat capacity and hence an unstable behaviour;
their correct thermodynamical interpretation is in terms of
sphalerons \cite{be,gpy}. The 2+1 black holes on the other hand have a positive
heat capacity which allows us to consider them a well-defined thermodynamical
system, characterized e.g. by a real partition function.

We will use the path integral approach to black hole thermodynamics, where
the following expression for the canonical partition function is used:
\be
Z(\beta)=\int {\cal D}g{\cal D}\phi\ e^{-I_{grav}[g]-I_{mat}[g,\phi]}. 
\label{pathint}\ee
$\beta$ is the period of the Euclidean time and is identified with
the inverse of the temperature. For the picture to be consistent, one has 
to integrate over periodic Euclidean metrics and matter configurations, the
actions $I_{grav}$ and $I_{mat}$ also being the Euclidean gravitational
and matter actions.
Going to imaginary time also fixes the topology of the Euclidean black hole 
manifold: generically in $d$ dimensions the Euclidean topology of a black 
hole is $R^2\times S^{d-2}$; the $R^2$ factor includes the possibility of 
a conical singularity.
This topology gives to the black hole a non vanishing horizon area, thus
disconnecting black hole solutions from (hot) flat space.

It is known that one can reproduce all the $n$-point functions of a
quantum field theory from the path integral of the action of a relativistic
particle (see for instance Ref.~\cite{polya}). 
To be more specific, one can take a particle of mass
$m$ and action: 
\be 
I_{part}=m L(path)=m \int ds 
= m \int \sqrt{g_{\mu \nu}\dot{x}^{\mu} \dot{x}^{\nu}}\ ds
\label{ipart}\ee
to mimic the quantum field theory of a real massive scalar field.
The vacuum fluctuations of the free field will be described by paths
of $S^1$ (loop) topology. These loops are necessarily off-shell in
the Lorentzian picture, so when going to imaginary time there are no
restrictions of causal type on them.
Hence, to calculate the partition function, one sums over all closed 
Euclidean paths of the relativistic particle. The expression \rref{pathint}
is thus equivalent to:
\be
Z(\beta)=\int {\cal D}g{\cal D}x(s)\ e^{-I_{grav}[g]-I_{part}[g,x(s)]}.
\label{pathint2}\ee

Now the problem of coupling particle matter to 2+1 gravity is also
well-known \cite{djt}: the only back reaction caused by the particles
is to create a conical singularity along their paths.
The main scope of this paper is to show, treating the
matter action as in \rref{pathint2}, that the matter indeed contributes
to the tree level entropy of the black hole, by a mechanism which includes
a dynamical conical singularity.
This contribution leads to a decrease 
of the purely gravitational value of the entropy, thus
correcting the law \rref{arealaw}.

The paper is organized as follows: in section 2 we find a classical solution to
Euclidean 2+1 gravity coupled to a relativistic particle; this solution
represents macroscopical topologically non-trivial fluctuations of the
matter field; we also derive
an equation of motion for the deficit angle. In section 3
we analyse the thermodynamics of the black hole with the dynamical
conical singularity and we derive the corrected area law for entropy.
In the concluding section we try to provide a link with results in
other dimensions \cite{ehw}, and we 
briefly speculate on how to consider at the same
time the different solutions corresponding to a different winding number.

\section{An equation for the deficit angle}

The action of Euclidean 2+1 gravity coupled to a relativistic particle is the
following:
\be I=I_{grav} + I_{part}. \label{itot} \ee

In the Euclidean section, the black hole has topology $R^2\times S^1$, 
including the possibility of a deficit angle in the $R^2$ factor.
As a consequence of this topology, there is a point in the $R^2$ factor
for which the $S^1$ circle is of minimal size. This submanifold is 
identified with (the analytic continuation to imaginary time of) the 
horizon.

Thus, the closed paths divide naturally into homotopy
classes, characterized by how many
times the path winds around the $S^1$ factor. The condition for a
trajectory to belong to a particular class is:
\be \Delta_{tot}\phi \equiv \int_{path}d\phi= 2\pi k;
\label{trajcond} \ee
$\phi$ is the variable labelling the $S^1$ factor; 
we allow for both signs of $k$.

The minimum of $I_{part}$ is clearly attained, for a particle
moving according to \rref{trajcond}, when the path is wound around the
minimal circle at the horizon;
homotopically disjoint solutions are obtained choosing different integer values
of $k$.
These trajectories seem tachyonic at a first look, but we have to
consider them on the same footing as any other since we are computing 
a partition function. They are macroscopic vacuum configurations
due to the non-trivial topology of the space-time 
of the (Euclidean) black hole.
It has also to be noted that these are the only closed 
path classical solutions to the geodesic equations, and this will be the
reason why they contribute to the partition function.

Once the matter equations of motion are solved, the matter
action \rref{ipart} takes the simple form:
\be I_{part} = m |k| L_h, \label{imonsh} \ee
where $L_h$ is the circumference of the horizon.

Let us now consider the gravitational action and the back-reaction of the
matter on the geometry. 
The gravitational action $I_{grav}$ is taken to be covariant and 
supplemented by local boundary terms at infinity:
\be
I_{grav} = - \frac{1}{16\pi G} \int d^3 x \sqrt{g} \left( {\cal R} +
\frac{2}{l^2} \right) + B_{\infty}.
\label{igrav} \ee
The cosmological constant $\Lambda = - 1/l^2$ is negative
to incorporate BTZ black hole solutions which are asymptotically
anti-de Sitter \cite{bhtz}.

An easy way to derive the equations of motion and the on-shell value of the 
gravitational action is to start from the hamiltonian
action, which is vanishing on shell for a static black hole.
For a static, circularly simmetric metric, which we take here for
simplicity, we have:
\be ds^2= N^2 dt^2 + F dr^2 + R^2 d\phi^2 \label{metric} \ee
and the hamiltonian action reduces to:
\be
I_{ham}=\int dt\ d^2x N {\cal H}, \label{iham}\ee
where ${\cal H}$ is the hamiltonian constraint.   

Now, to render this action covariant at the horizon (i.e. to avoid
a `false' boundary term at the horizon in the variation of the action), 
one has to add a local
boundary term there \cite{btz94}.
This term turns out to be $-\frac{1}{4G}L_h$. 

When turning to analyze the boundary term at infinity in the variation
of $I_{ham}$, we find that the equations of motion are obtained when
$\delta M$ is put to $0$, $M$ being the ADM mass of an asymptotically
anti-de Sitter space-time, as defined\footnote{
A subtraction term is needed in the definition. We
choose the vacuum black hole
(i.e. the solution corresponding to $M=0$) as the reference space-time.}
in \cite{bcm}. One then calls this covariant action $I_{micro}$ with
reference to the thermodynamical microcanonical ensemble in which the
energy of the system is fixed \cite{by}:
\be 
I_{micro}=I_{ham}-\frac{1}{4G}L_h.
\label{iiham}\ee

However, it is
useful to have an action suitable for fixing $\beta$,
the periodicity of the Euclidean time $t$, instead of $M$. As
these two variables are canonically conjugated \cite{bhtz},
it is straightforward to obtain
it:
\be I_{can}=I_{micro}+\beta M, \label{ican}\ee
this being the canonical action since now the equations of motion are
obtained by fixing $\beta$, i.e. by fixing the temperature as in the 
canonical ensemble. Note that $\beta M$ is a local boundary term at infinity.

We turn now to solving the equations of motion for the metric.
The bulk of the variation of the gravitational action yields simply the 
equations of pure 2+1 gravity with a negative cosmological constant.
We pick the BTZ black hole solution, which is \rref{metric} with:
\be
N^2=F^{-1}=\frac{r^2}{l^2}-8GM, \quad \quad R=r.
\label{btzmetric}\ee

The `false' boundary term at the horizon of the variation of $I_{grav}$, which
is the same for $I_{micro}$ and $I_{can}$, is \cite{btz94}:
\be \delta I_{grav}|_{hor} = \frac{1}{4G} \left( \frac{\beta}{\beta_H}-1
\right)\delta L_h, \label{horvar} \ee
where $\beta_H$ is the inverse Hawking temperature, a function
of the metric parameters (say $M$):
\be \frac{2\pi}{\beta_H}=\left. \frac{N'}{\sqrt{F}}\right|_{r=r_h}\equiv
\kappa \label{betah}
\ee
and $L_h=2\pi R_h$ is the size of
the horizon.

Once the matter equations of motion are solved,
the variation of the matter action \rref{imonsh} with respect to the
gravitational degrees of freedom  reads:
\be \delta_{(grav)} I_{part}= m |k| \delta L_h. \label{matvar} \ee

Combining \rref{horvar} and \rref{matvar} and the requirement that
$L_h$ should not be fixed in advance (i.e. demanding covariance),
the condition $\delta I=0$ yields, for the horizon contribution, the
equation of motion for the (dynamical) deficit angle:
\be \left( 1- \frac{\beta}{\beta_H}\right)=4G|k|m. \label{dda2}\ee
We can rearrange this equation to obtain a condition on the periodicity
$\beta$:
\be \beta=\beta_H (1-4G|k|m). \label{dda} \ee

This is the only effect of the matter on the geometry. Notice that $\cal R$
is then singular at $r=r_h$, but this singularity is generated by the
matter distribution, so no conceptual problem is arising.

Note that we could have straightforwardly inserted
the stress--energy tensor of the relativistic particle
into the Einstein
equations finding a conical singularity along the particle's path, as in
the work of Deser, Jackiw and 't Hooft \cite{djt}.
We preferred instead to give an alternative
derivation in a formalism that yields directly an equation for the
deficit angle, which measures the strength of the conical singularity.

We will not consider the case for which $\beta<0$, i.e. when the deficit angle 
exceeds $2\pi$, because it is unphysical when only one particle is
present \cite{djt}. Also, the case for which $\beta=0$ is a degenerate one
because the topology turns from the cone to the cylinder, ceasing to be
a black hole topology; we will not
consider it either.

We now turn to the thermodynamics, which can be derived
either from $I_{micro}$ or from $I_{can}$.

\section{The thermodynamics}

To describe the thermodynamical behaviour of 2+1 black holes we use
the path integral representation of the partition function. In this
formalism, the black hole thermodynamics has a clear quantum mechanical origin
(i.e. the $\hbar$ dependence of the quantities is naturally fixed).
We consider in the following the semiclassical approximation
in which we retain only the leading order of a saddle point
(tree level approximation).

Let us first calculate the temperature\footnote{
One has to spend a few words on the
thermodynamical conjugate variables $\beta$ and $M$.
Since the geometry is asymptotically anti-de Sitter, these two
quantities do not coincide with the local inverse temperature and the energy at
infinity. This is due to the fact
that the Killing vector $\frac{\partial}{\partial t}$
has an infinite norm at infinity. Rather, $\beta$ can be thought of as the
global inverse 
temperature; its thermodynamical conjugate is $M$ and can moreover be
identified with the ADM mass \cite{bhtz,bcm}.}
of the solution found in the preceding section.
For a BTZ black hole, eqs.~\rref{betah} and \rref{dda} give:
\be \beta=\frac{\pi l}{\sqrt{2GM}}(1-|k|\eta),\quad \eta=4Gm. \label{betadda}
\ee

This relation entails that, for $|k|\eta<1$ (the physically relevant
case), the temperature
$T\sim \sqrt{M}$. This implies that the BTZ black hole has a positive
heat capacity, because $c_V=\frac{\partial M}{\partial T}\sim T >0$. This
behaviour is in sharp contrast with the case of the 4-dimensional Schwarzschild
black hole, which has a negative heat capacity and hence an unstable
thermodynamic behaviour. In our case, the positivity of $c_V$ will
allow us to use a well defined (real) canonical partition function.

The standard way of calculating the canonical partition function
is to evaluate \rref{pathint} taking $I_{grav}$ to be $I_{can}$: we integrate
over all Euclidean metrics of fixed period $\beta$ \cite{gh}.

There is an alternative derivation \cite{by} (see also \cite{btz94})
that uses $I_{micro}$ instead of $I_{can}$. 
The path integral using this action
yields directly the exponential of the entropy, i.e. the microcanonical
partition function or the density of states:
\be e^{S(M)}=\int {\cal D}g{\cal D}x(s)\ e^{-I_{micro}[g]-I_{part}[g,x(s)]}. 
\label{pathintmicro}\ee
We integrate over the Euclidean metrics which are periodic in the
Euclidean time and for which the ADM mass is fixed to $M$.

Evaluated by the saddle point approximation to the leading order and for a
fixed value of the winding number $k$, \rref{pathint} gives:
\begin{eqnarray} -\ln Z_k(\beta)&\simeq& 
(I_{can}+I_{part})|_{on-shell}\nonumber \\
&=&\beta M -\frac{L_h}{4G}+ m|k|L_h \nonumber \\
&=& \beta M -
\frac{L_h}{4G}(1-|k|\eta). \label{pfdda} \end{eqnarray}

One has to be careful at this stage because $M$ and $L_h$, as given by
the ADM mass formula and by $L_h=2 \pi R_h$,
are functions of $\beta_H$ rather than $\beta$; when
this is taken into account, $\ln Z_k(\beta)$ has a nonlinear dependence
on the winding number $k$:
\be \ln Z_k(\beta)=\frac{\pi^2 l^2}{2G\beta}(1-|k|\eta)^2. \label{pfdda2}\ee

The entropy is then obtained from the standard thermodynamical relation:
\be S(M)= - \left.\left(\beta\frac{\partial}{\partial\beta}-1\right)
\ln Z(\beta) \right|_{\beta=\beta(M)},
\label{entr} \ee
where $\beta=\beta(M)$ is obtained by inverting the
equation $M=-\frac{\partial}{\partial\beta}\ln Z(\beta)$. The result is
thus:
\be S_k=\frac{L_h}{4G}(1-|k|\eta)=\pi l\sqrt{\frac{2M}{G}}(1-|k|\eta).
\label{entrdda} \ee

This is exactly the same result that we could have obtained by use of
\rref{pathintmicro} and \rref{iiham}, a consequence of the equivalence
of the two approaches when the contribution of only one saddle point is
taken into account.
Since $S_k$ is a function of $M$, it has a linear dependence on $k$.

The effect of the dynamical conical singularity on the thermodynamics
of the black hole is hence the following: with respect to the $k=0$ case,
the entropy is lowered and the free energy $F_k (\beta)=-\beta^{-1} \ln
Z_k (\beta)$ is increased. Hence, the black hole topology allows for
nontrivial configurations of the matter fields, here represented by
a relativistic particle, which lead to a correction of the law 
\rref{arealaw} for black hole entropy already at tree level.

As a concluding remark on the thermodynamical behaviour of our 
solution, notice that
if we want $Z_k$ to be represented by a Laplace transform of the
density of states $e^{S_k}$,
we need a condition on the temperature.
Indeed, the integral:
\be Z_k(\beta)=\int_0^{\infty}dM\ e^{-\beta M} e^{S_k(M)} \label{lt}\ee
can be calculated by saddle point approximation only if the following
condition holds, assuring that the integral can be approximated by a
gaussian integral:
\be  \left|\frac{\partial^2 S_k}{\partial M^2}\right|^{
-\frac{1}{2}}_{M=M_{s.p.}}\ll M_{s.p.}, \label{spv}\ee
where $M_{s.p.}$ is the value of $M$ at the saddle point, expressed in
terms of $\beta$. The condition \rref{spv} becomes then:
\be \beta\ll \frac{\pi^2 l^2}{4G}(1-|k|\eta)^2, \label{hightemp}\ee
which means high temperature.

\section{Concluding remarks}

Let us first put forward the conclusions of the preceding sections.
We have shown that a 2+1 black hole with a particle
path wound around its horizon violates the (three-dimensional) standard
area law for its entropy.
Indeed, in the presence of a particle the topological $R^2$ factor of the
black hole manifold has in fact the shape of a cone near the horizon.
The entropy of this black hole is simply proportional to the total
opening angle.

This result is very similar to the 3+1 dimensional computation of
\cite{ehw}, independently of the different thermodynamics of BTZ and
Schwarzschild black holes. 
However, the physical interpretation behind the effect is slightly
different.

In 2+1 dimensions, the worldline of the particle is the 
representation of a macroscopic vacuum fluctuation. It is on this ground
that is has to be included in the computation of the partition function.
On the other hand, in 3+1 dimensions one is faced with a string
configuration whose worldsheet has the topology of a sphere, which
does not represent in (closed) string theory conventional vacuum fluctuations;
hence, in this case one is handling a more complicated phenomenon.

One could try to extend this result to dimensions higher than 4.
Generically, one can guess that the same effect can arise for
a $d$ dimensional black hole when the Euclidean worldvolume of
a $(d-3)-$brane, of $S^{d-2}$ topology, is wrapped around the
horizon. Still, the physical interpretation of such a worldvolume
should be made clear.

Going back to the 2+1 dimensional case, one can make some
further considerations.

Whether the result obtained is a true violation of the area law \rref{arealaw}
or not depends on the stability
of a solution at fixed winding number $k$. If the lifetime of such a solution
is long enough to establish a thermodynamic equilibrium, then
the entropy of this configuration has a physical significance. If on the other
hand it appears that this solution is highly unstable and decays rapidly
to another configuration, then
this result is not of great relevance. Indeed, the $k=0$ solution is the
one of the lowest action\footnote{Also compared to the vacuum black hole, which
has vanishing on-shell action and arbitrary temperature.} and free energy.

These last considerations lead us to considering the more general problem
where all the solutions are taken into account, i.e. when there are no
more restrictions on the winding number $k$.
Since a solution exists for each winding number $k$, then in the path
integral \rref{pathint} or \rref{pathintmicro} the integrand has
several minima. One is tempted to sum over them, evaluating each one
by a saddle point approximation \cite{cpw,dgt}.

However, since our aim is to calculate thermodynamical averages and
not quantum expectation values, this procedure is not straightforward.
Indeed, all the standard thermodynamical relations are retrieved from the
saddle point approximation of the Laplace transform \rref{lt} or its
inverse. There is no obvious reason for which this procedure still should
work when the
partition function or the density of states are derived from a path
integral consisting of several saddle points.
In particular, the microcanonical and
the canonical approaches appear inconsistent in the general case.
This problem, which goes beyond the framework of 2+1-dimensional
black holes, is still under consideration.

\subsection*{Acknowledgments}

I would like to thank Fran\c{c}ois Englert for important and helpful
remarks and discussions
during the realization of this work. I also benefited from interesting
discussions with Marc Henneaux, Laurent Houart, Paul Windey and
especially Peter van~Driel.

\end{document}